\documentclass{eds2019}

\usepackage{amsmath}
\usepackage{breqn}
\usepackage{amsfonts}
\usepackage{dsfont}
\def\be{\begin{equation}}
\def\ee{\end{equation}}
\def\bea{\begin{eqnarray}}
\def\eea{\end{eqnarray}}

\newcommand{\Photo}%{\includegraphics[height=35mm]{mypicture}}

\begin{document}
\vspace*{4cm}
\title{SCHWINGER-BASED QCD  FORMULATION'S DERIVATION OF ELASTIC PP SCATTERING}

\author{H.M. Fried, Y.M.Sheu, P.H.Tsang\footnote{speaker}}

\address{Brown University, Department of Physics, RI, USA}

\author{Y. Gabellini, T.Grandou}
\address{Institut de Physique de Nice, UMR 7010 CNRS, Site Sophia 06560 Valbonne, France}

\maketitle\abstracts{Using previously described functional techniques for some non-perturbative, gauge invariant, renormalized QCD processes, a simplified version of the amplitudes - in which forms akin to Pomerons naturally appear - provides fits to ISR and LHC-TOTEM pp elastic scattering data. Those amplitudes rely on a specific function $\phi(b)$ which describes the fluctuations of the transverse position of quarks inside hadrons.}

\section{Introduction}

This talk is covers the work in ~\cite{isrlhc}. Beginning with Schwinger's Generating Functional and the QCD Lagrangian, applying two procedures that were overlooked in the last four decades, an analytic, gauge-invariant correlation functions for Non-Perturbative QCD is obtained. 
This formulation produces the following results thus far:

\begin{itemize}
    \item First-principled dynamical quark confining potential for quarks \cite{fried2012}, 
    \item A potential obtained from QCD that allows nucleons to be bounded, thus provided the first-principled model deuteron \cite{fried2013},
    
    \item New property of Effective Locality, provides gauge-invariant summation of all gluonic exchanges between quarks; more over, the interaction becomes a local interaction. \cite{grandou2017}.
    \item Obtained Chiral Symmetry Breaking for dynamical quarks out of the new property of Effective Locality \cite{grandou2019}.

    \item Extended Asymptotic freedom as supported by other non-perturbative approaches: Dyson Schwinger Equation \cite{friedgrandou2012}.
    \item A qualitative description of the Hadron Confinement mass scale(s) \cite{fried2015}.
    \item The full SU(3) algebraic content of QCD amplitudes, both $C_2$ and $C_3$ casimir invariants are preserved \cite{grandou2014}.
    \item First-principled calculation of elastic proton-proton scattering at ISR and LHC energies. This will be the focus of this talk \cite{isrlhc}.
    
\end{itemize}

\subsection{Schwinger Generating Functional}\label{subsec:prod}

The Schwinger Generating Functional can be rewritten into gaussian operations on gaussian fields.

\begin{equation}
    \mathfrak{Z}_{QCD}[\bar{\eta},\eta,j] = \mathfrak{N} e^{-\frac{i}{2} \int \frac{\partial}{\partial A}\cdot D_F^{(0)}\cdot \frac{\partial}{\partial A}}\cdot e^{\frac{i}{4} \int \bf{F}^2 + \frac{i}{2} \int A\cdot(-\partial^2)\cdot A} \cdot e^{i\int \bar{\eta}\cdot \bf{G}_F[A]\cdot \eta+ \bf{L}[A]}|_{A=\int D_F^{(0)}\cdot j}
\end{equation}

where $G_F(x,y|A) = [m + \gamma\cdot(\delta - igA\tau)]^{-1}$, and $L[A]=ln[1-i\gamma A\tau[0]]$. 

The $\bf{F}^2$ can be linearized with Halpern's half a century old expression:

\begin{equation}e^{-\frac{i}{4} \int \bf{F}^2}=N \int d[\chi] e^{\frac{i}{4} \int \chi^2 + \frac{i}{2} \int F\cdot \chi} 
\scriptstyle
\end{equation}
where $\chi^a_{\mu\nu} = -\chi^a_{\nu\mu}$ is anti symmetric tensor.
It is this added $\chi$ field that plays the central role in the summation of all gluons in the non-perturbative regime.

\subsection{Gauge-Invariance}

Schwinger's Generating Functional now becomes 
\begin{equation}
        \mathfrak{Z}_{QCD}[\bar{\eta},\eta,j]=N \int d[\chi] e^{\frac{i}{4} \int \chi^2 } e^{\mathfrak{D}_A^{(0)}}\cdot e^{\frac{i}{2}\int \chi \cdot \bf{F} + \frac{i}{2} \int A\cdot (-\partial)\cdot A} e^{i\int \bar{\eta}\cdot G_F[A]\cdot \eta + L[A]}|_{A=\int D_F^{(0)}\cdot j}
\end{equation}

Calculating 2n-point Fermionic Green's functions (e.g. n=2), gives 
\begin{equation}
    =N\int d[\chi]e^{\frac{i}{4}\int \chi^2} e^{\mathfrak{D}_A^{(0)}}e^{\frac{i}{2}\int A \cdot (D_F^{(0)})^{-1}\cdot A}
    G_F(1|g A)G_F(2|g A)e^{L[A]}|_{A=0}
\end{equation}
\begin{equation}
    e^{\mathfrak{D_a}}F_1[A] =
    exp[\frac{i}{2}\int \bar{Q}\cdot D_F^{(0)}\cdot (1-\bar{K}\cdot D_F^{(0)})^{-1}\cdot \bar{Q}- \frac{1}{2} Tr \ln(1-D_F\cdot \tilde{K}]
    \cdot exp[\frac{1}{2}\int A\cdot \bar{K}\cdot (1-D_F^{(0)}\cdot \bar{K})^{-1}\cdot A]
\end{equation}

where 
\begin{equation}
    D_F^{(0)}\cdot (1- \bar{K}\cdot D_F^{(0)})^{-1} = D_F^{(0)}\cdot [1-\hat{K}+(D_F^{(0)})^{-1})\cdot D_F^{(0)}]^{-1}
    =-(\tilde{K}_{\mu\nu}^{ab}+gf^{abc}\chi_{\mu\nu}^c)^{-1}=-\hat{\bf{K}}^{-1}
\end{equation}
with 
$F_1[A]=e^{\frac{i}{2}\int A\cdot \bar{K}\cdot A+i\int\bar{Q}\cdot A}$, 
$F_2[A] = e^{L[A]}$
and $<z|\bar{K}_{\mu\nu}^{ab}|z'>=[\tilde{\bf{K}}^{ab}_{\mu\nu}(z)+gf^{abc}\chi_{\mu\nu}^c(z)]\delta^{(4)}(z-z')+<z|(D_F^{(0)})^{-1}|_{\mu\nu}^{ab}|z'>$.

We then have 
\begin{dmath}
    e^{\mathfrak{D}_A}F_1[A]F_2[A]= exp[-\frac{i}{2}\int \bar{Q}\cdot \hat{K}^{-1}\cdot \bar{Q}]+ \frac{1}{2}Tr\ln \hat{\bf{K}} + \frac{1}{2} Tr \ln(-D_F^{(0)})] \\ 
    \cdot exp[\frac{i}{2}\int \frac{\partial}{\partial A}\cdot D_F^{(0)}\cdot \frac{\partial}{\partial A'}]\cdot exp[\frac{i}{2}\int \frac{\partial}{\partial A'} \cdot \hat{\bf{K}}^{-1}\cdot \frac{\partial}{\partial A'} -\int \bar{Q}\cdot \hat{\bf{K}}^{-1}\cdot \frac{\partial}{\partial A'}]\cdot (e^{\mathfrak{D}_A} F_2[A'])
\end{dmath}
and
\begin{dmath}
    e^{\mathfrak{D}_A}F_1[A]F_2[A] = N exp[-\frac{i}{2}\int \bar{Q}\cdot \hat{\bf{K}}^{-1}\cdot \bar{Q}+\frac{1}{2}Tr \ln\hat{\bf{K}}]\cdot exp[\frac{i}{2}\int \frac{\partial}{\partial A}\cdot \hat{\bf{K}}^{-1}\cdot \frac{\partial}{\partial A} - \int \bar{Q}\cdot \hat{\bf{K}}^{-1}\cdot \frac{\partial}{\partial A}]\cdot exp(L[A])|_{A\rightarrow 0}
\end{dmath}

From above, all explicit gauge dependencies are cancelled. That is, Gauge Invariance is explicitly preserved by means of Gauge Independence.

$-\bf{\hat{K}}^{-1}\equiv$$ (f \cdot \chi)^{-1}$ represents all gluonic exchanges between quarks, a $\textbf{Gluon\  Bundle}$, ie, all gluonic exchanges summed in the non-perturbative regime.

 In QCD, where confinement and chiral symmetry breaking hold, the impact parameter, $b$ must fluctuate \cite{casher_79} \cite{brodsky_shrock_09}. At this stage, we choose a fluctuating $b$ by a deformed gaussian (gaussian resulted in zero potentials). 
 \begin{equation}
     \varphi(b)=\frac{\mu^2}{\pi}\frac{1-\xi/2}{\Gamma(\frac{1}{1-\xi/2})}e^{-(\mu b)^{(2- \xi)}}
 \end{equation} where $0<\xi<<1$

With this, non-perturbative QCD processes becomes processes of gluon bundles, $(f\cdot \chi)^{-1}$'s and closed quark loops, $L[A]$'s.
With the introduction of a particular renormalization scheme as described in \cite{fried_et_al_2015}, functional integrals of Halpern's $\chi$'s reduces to ordinary integrals, where $ d\chi^4$ becomes space-time integrals of infinitesimal sizes, giving $\delta$ functions joining with closed quark loops, $L[A]$'s.

\section{Comparing theory with experimental proton-proton elastic scattering differential cross section at ISR and LHC energies.}

\begin{figure}
\begin{minipage}{0.5\linewidth}
\centerline{\includegraphics[width=1.0\linewidth]{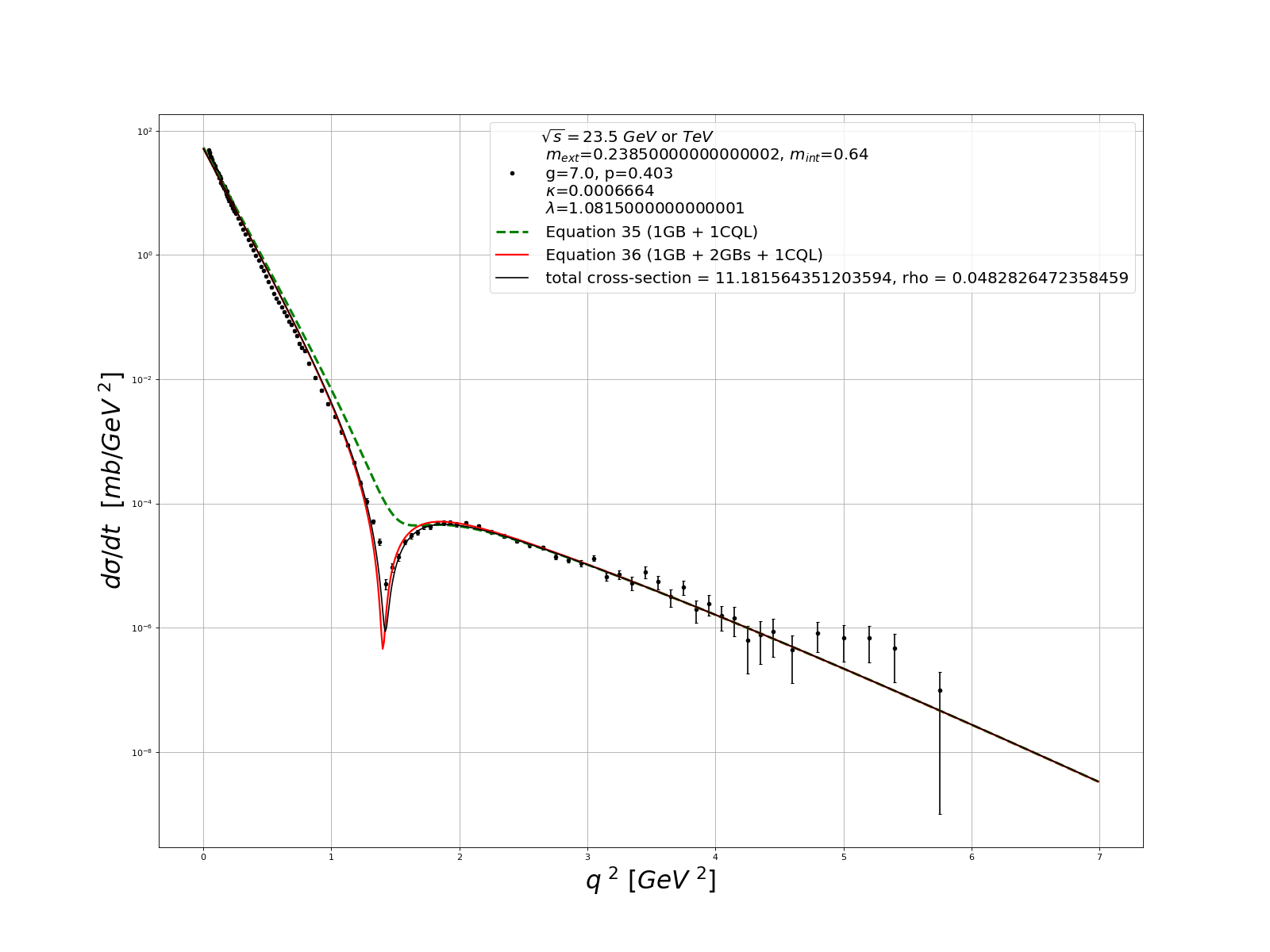}}
\caption{ISR = 23.5 GeV}
\end{minipage}
\hfill
\begin{minipage}{0.5\linewidth}
\centerline{\includegraphics[width=1.0\linewidth]{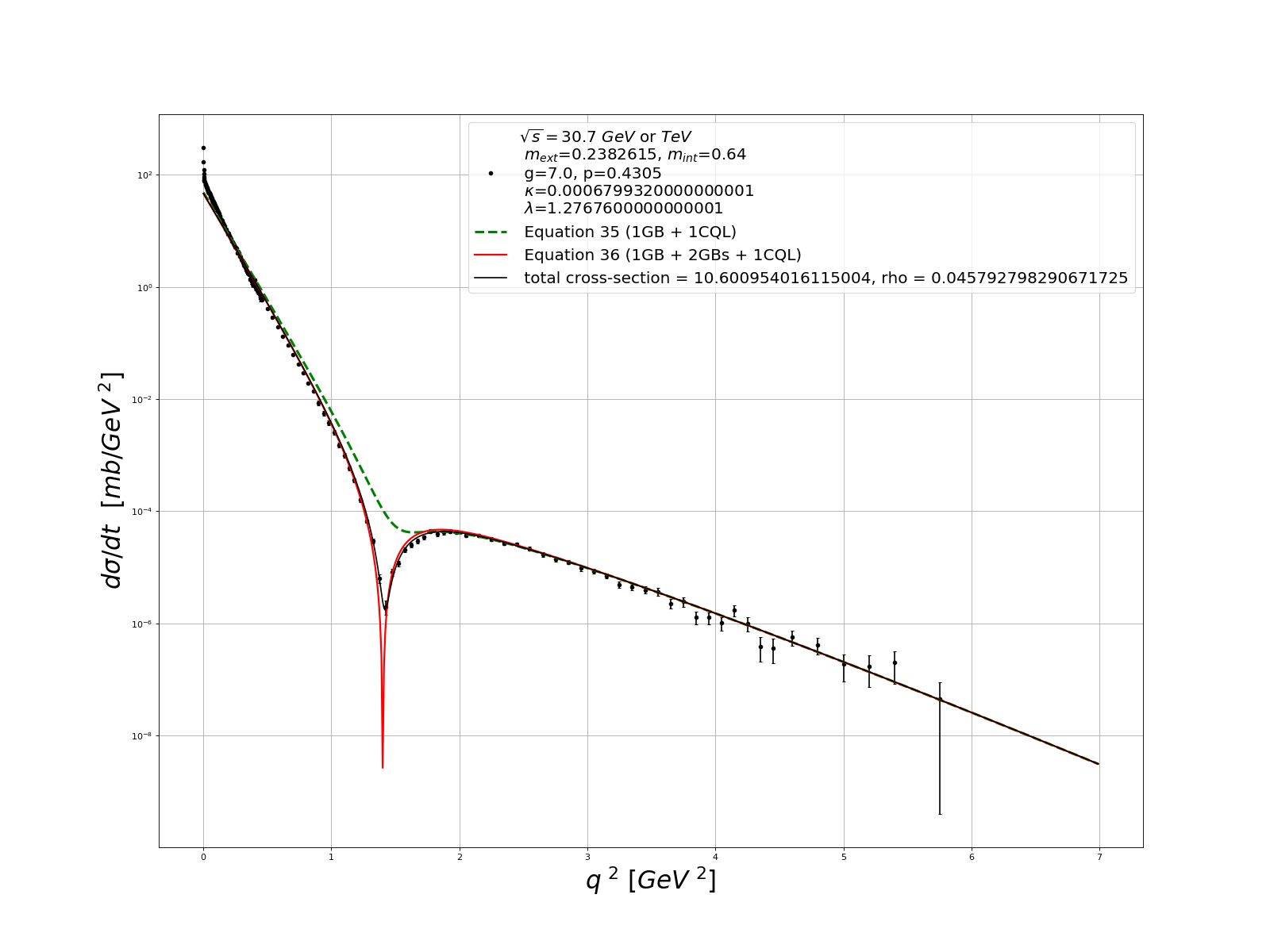}}
\caption{ISR = 30.7 GeV}
\end{minipage}
\hfill
\begin{minipage}{0.5\linewidth}
\centerline{\includegraphics[angle=0,width=1.0\linewidth]{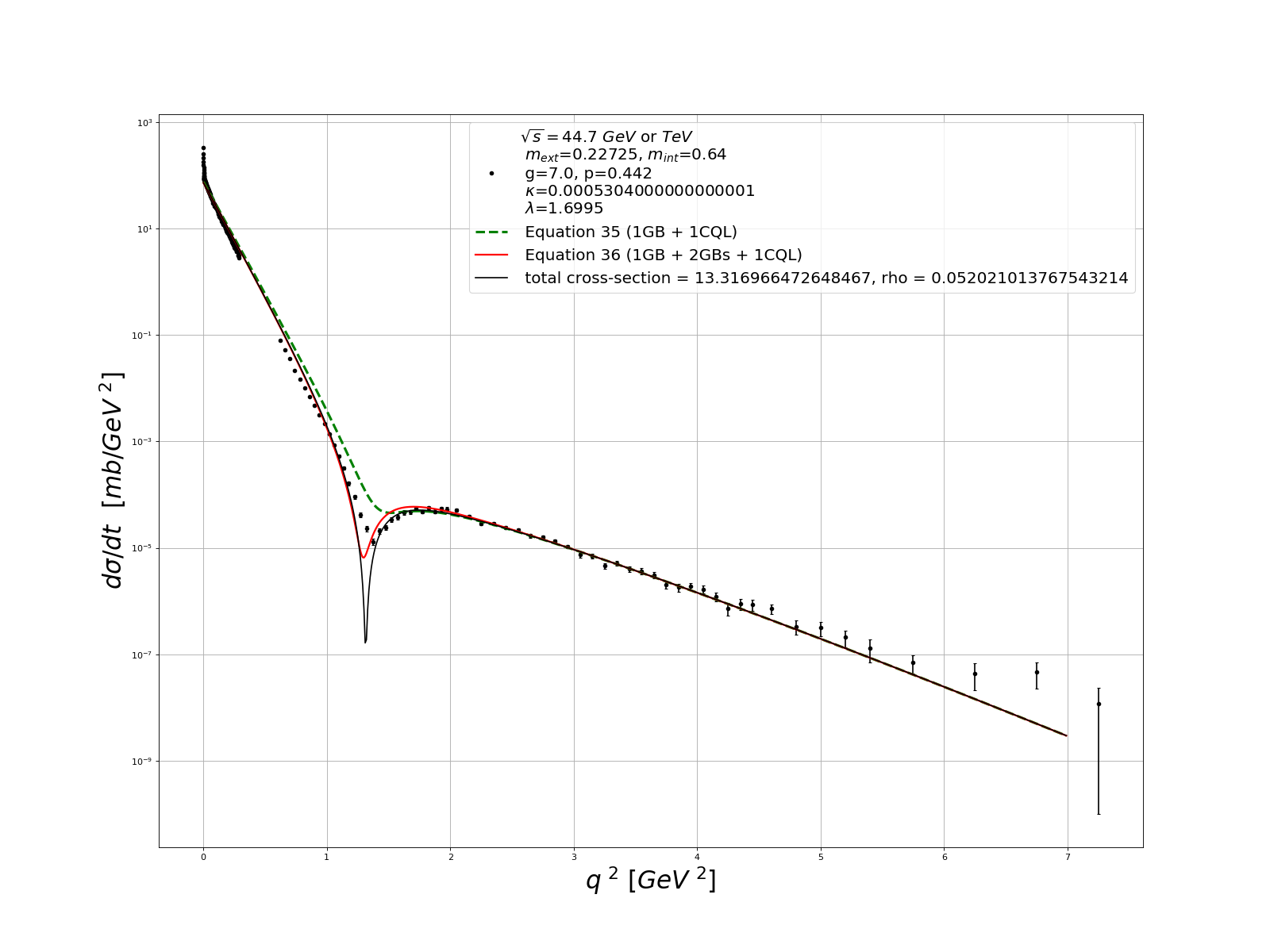}}
\caption{ISR = 44.7 GeV}
\end{minipage}
\begin{minipage}{0.5\linewidth}
\centerline{\includegraphics[angle=0,width=1.0\linewidth]{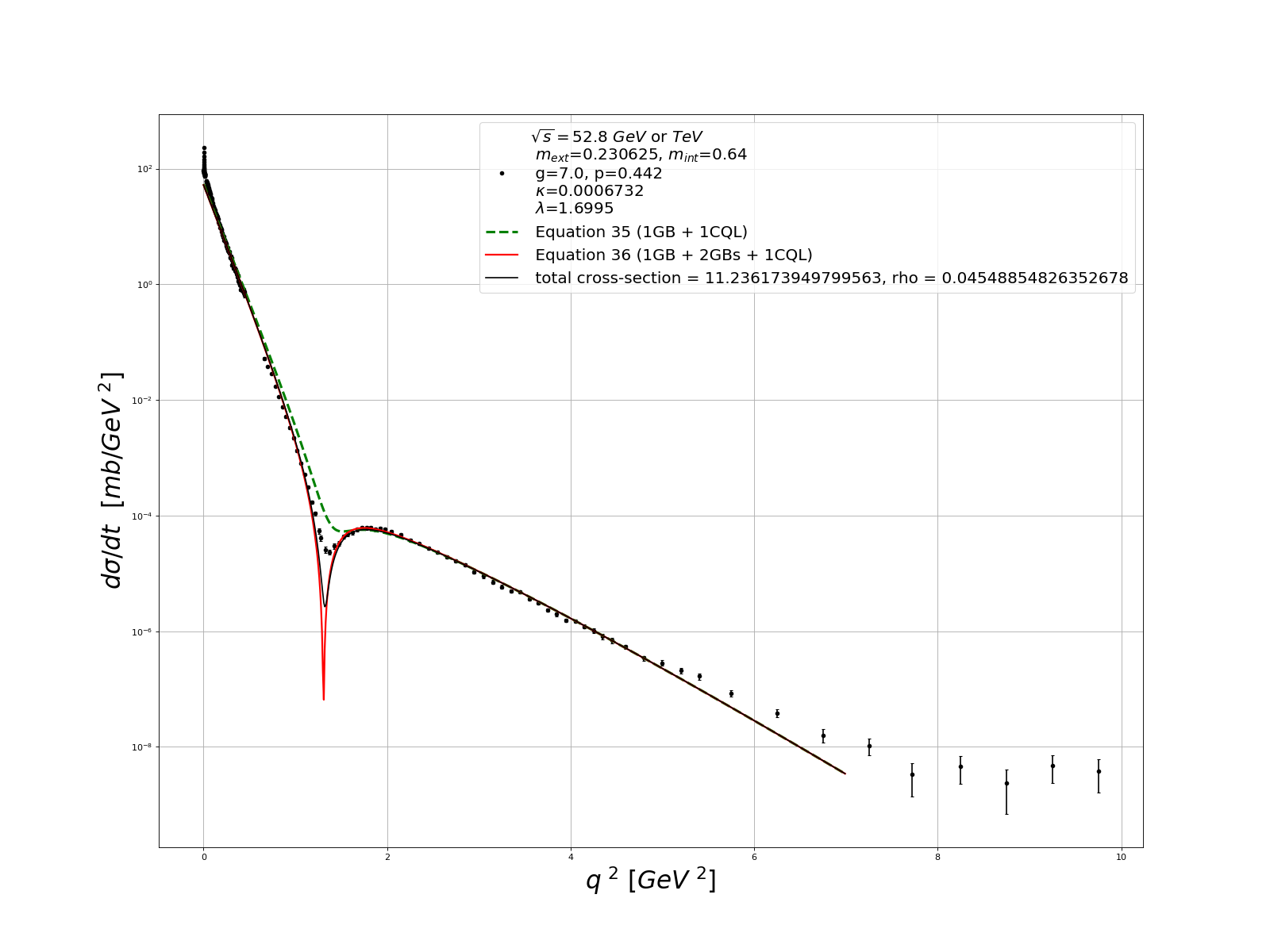}}
\caption{ISR = 52.8 GeV}
\end{minipage}
\begin{minipage}{0.5\linewidth}
\centerline{\includegraphics[angle=0,width=1.0\linewidth]{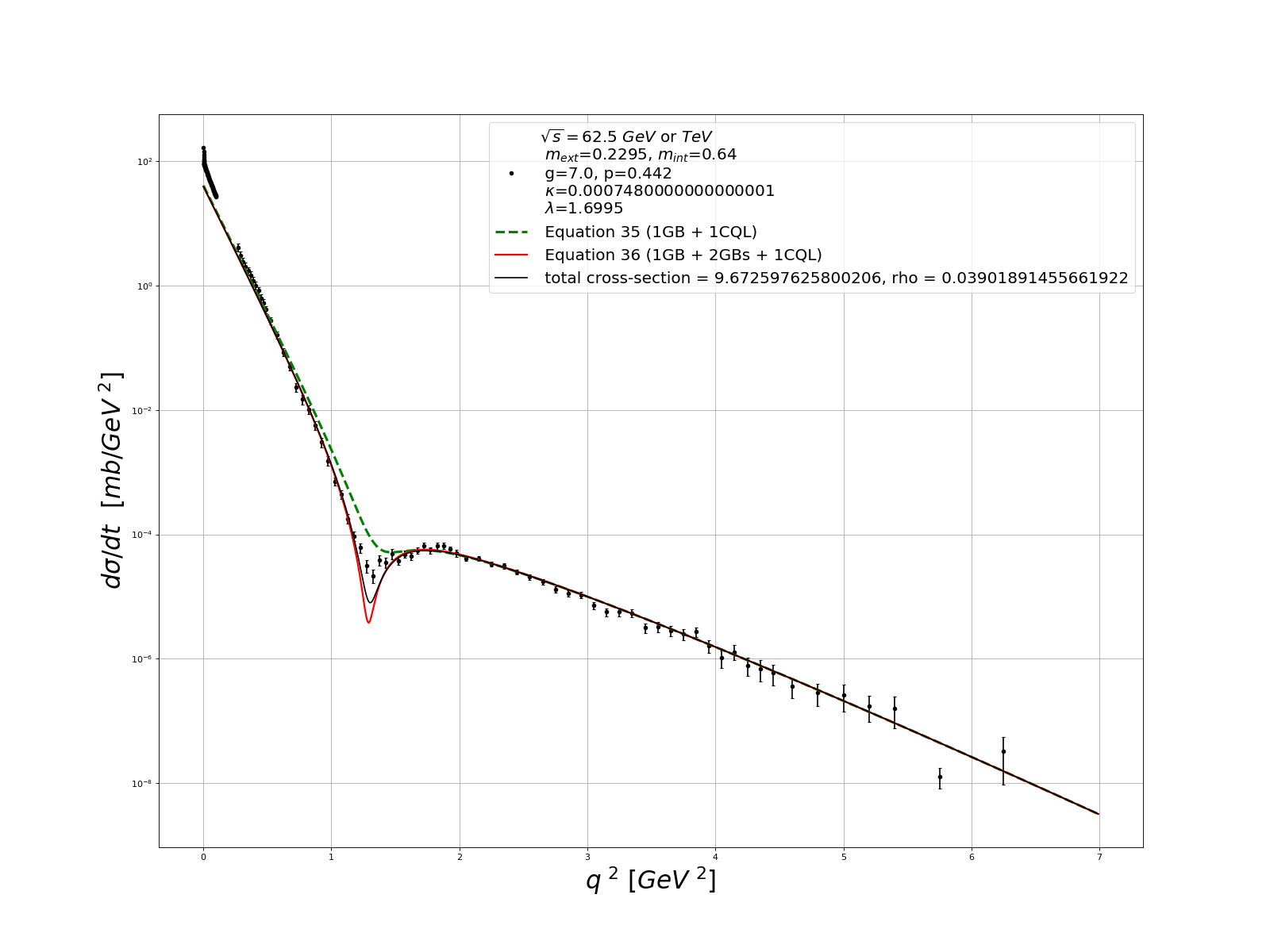}}
\caption{ISR = 62.5 GeV}
\end{minipage}
\caption[]{ISR energies for elastic proton-proton scattering. Green dotted line for one Gluon Bundle and one Quark Loop Chain. Red solid line for One Gluon Bundle plus two Gluon Bundles and one Quark Loop Chain.}
\end{figure}

\begin{figure}
\begin{minipage}{0.5\linewidth}
\centerline{\includegraphics[angle=0,width=1.0\linewidth]{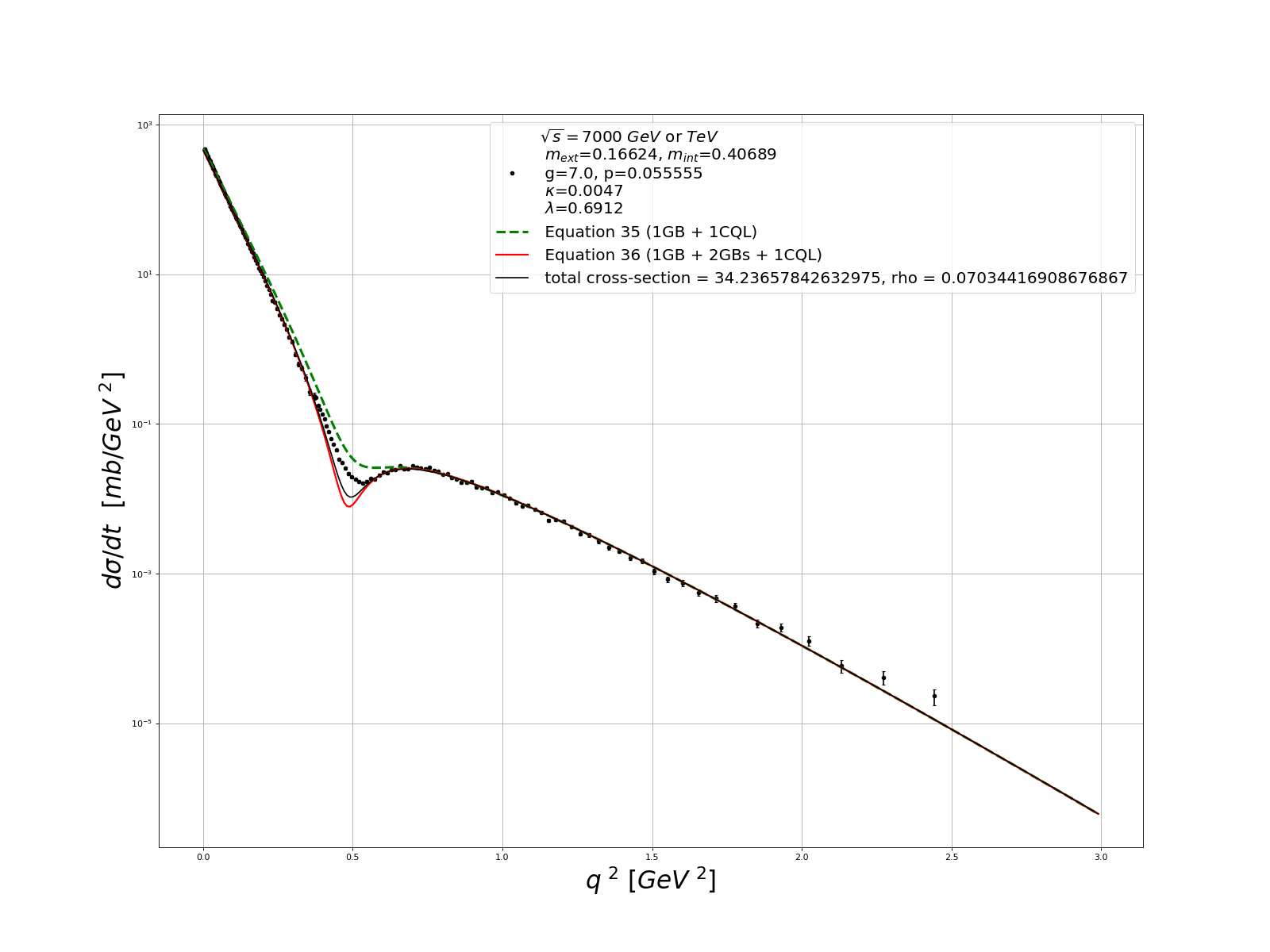}}
\caption{LHC-TOTEM = 7.0 TeV}
\end{minipage}
\begin{minipage}{0.5\linewidth}
\centerline{\includegraphics[angle=0,width=1.0\linewidth]{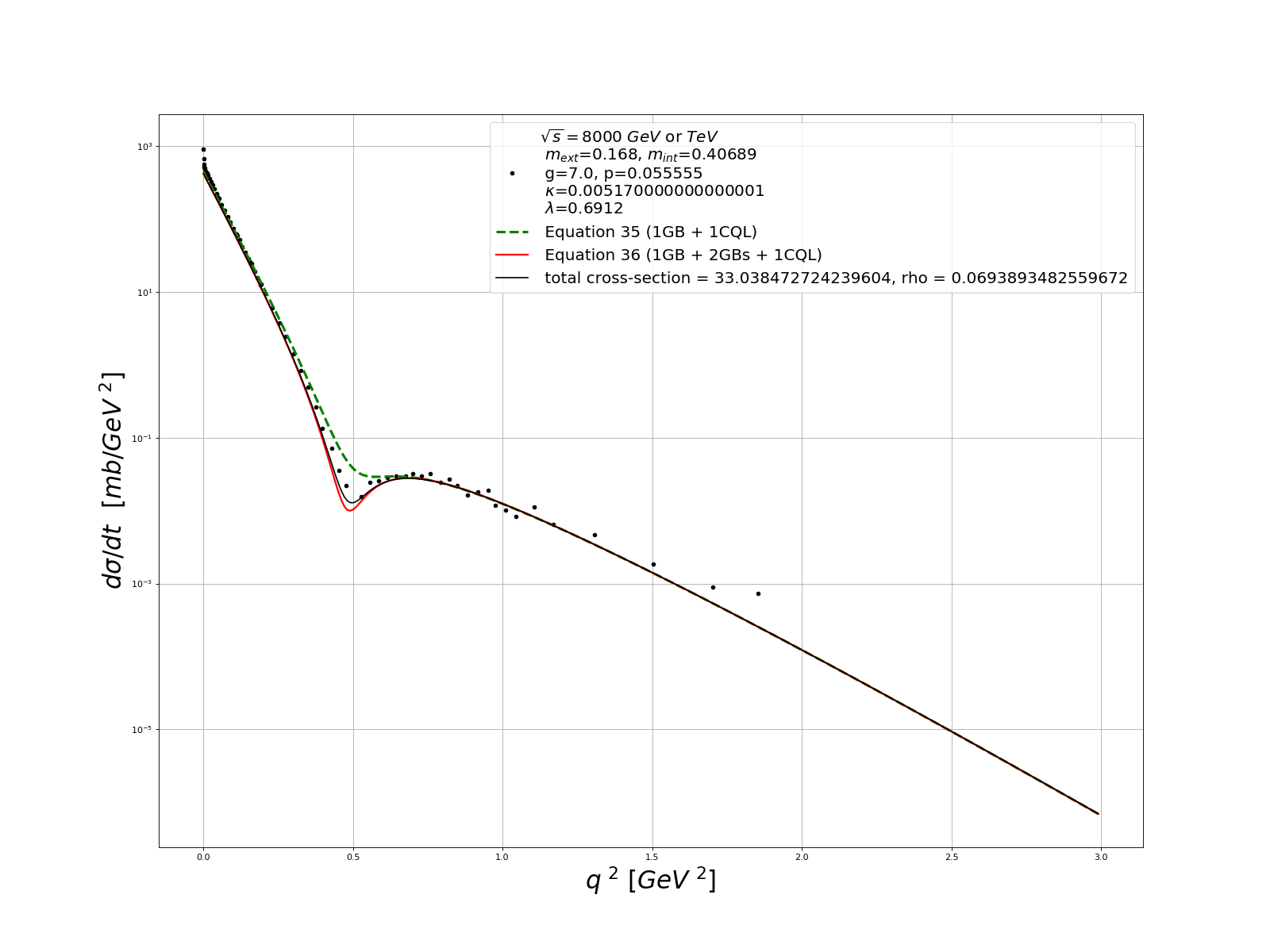}}
\caption{LHC-TOTEM = 8.0 TeV}
\end{minipage}
\begin{minipage}{0.5\linewidth}
\centerline{\includegraphics[angle=0,width=1.0\linewidth]{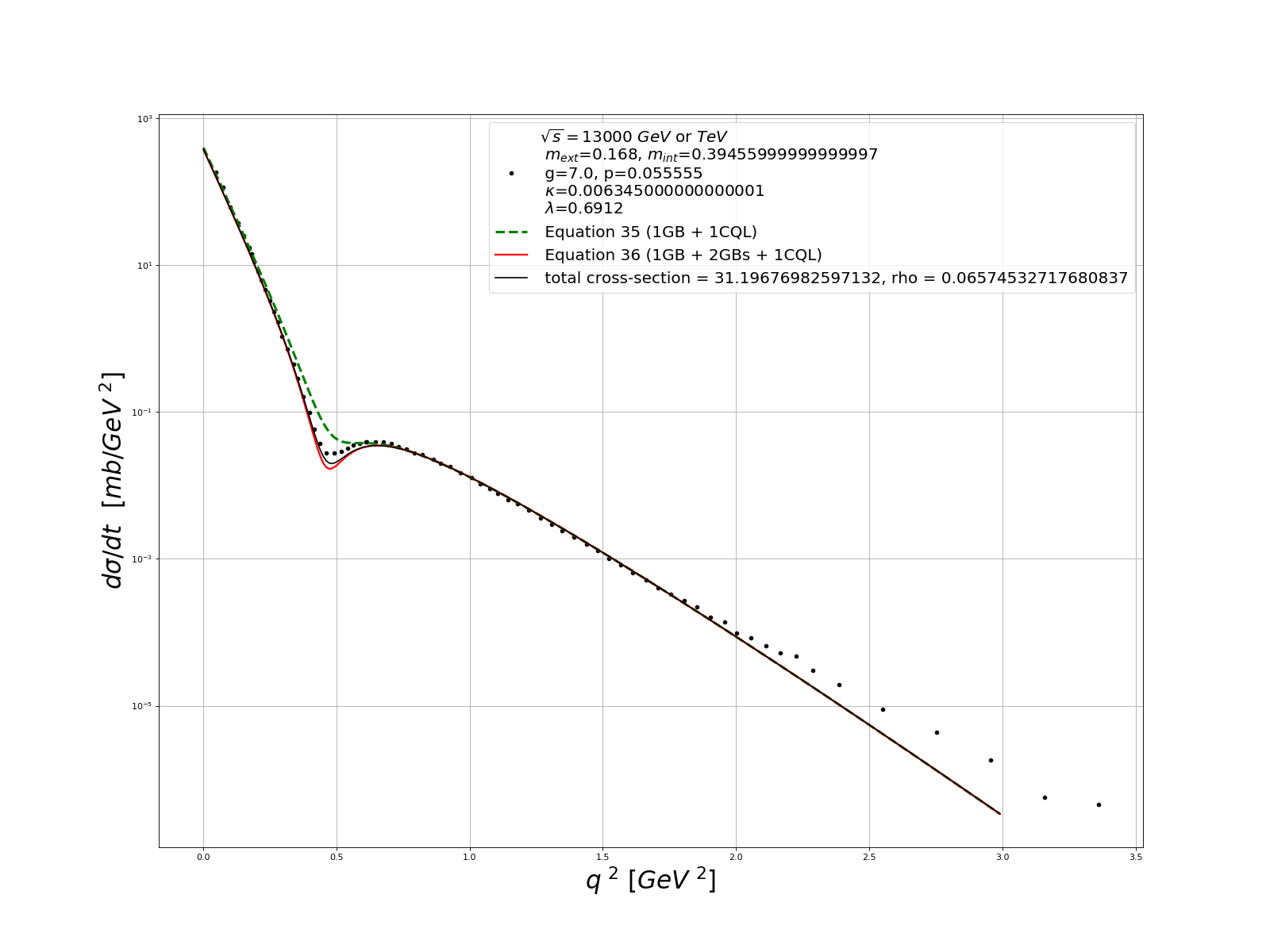}}
\caption{LHC-TOTEM = 13.0 TeV}
\end{minipage}
\caption[]{LHC-TOTEM energies for elastic proton-proton scattering. Green dotted line for one Gluon Bundle and one Quark Loop Chain. Red solid line for One Gluon Bundle plus two Gluon Bundles and one Quark Loop Chain.}
\label{fig:radish}
\end{figure}

As shown in \cite{isrlhc}, \cite{fried_et_al_2015}, the choice for $\delta^2 \ell = \kappa / \bar{m} $ results explicit $0$ for all graphs except Quark Loop Chain graphs.

The non-perturbative QCD scattering amplitude can thus be calculated as 
\begin{equation}
    T(s,\vec{q})= \frac{is}{2M^2}\int d^2 b e^{i\vec{q}\cdot\vec{b}}[1-e^{i \mathds{X}^{pp}(s,\vec{b})}]
\end{equation}
Where $\mathds{X}^{pp}$ is the Eikonal of the proton-proton process where $e^{i\mathds{X}^{pp}}=e^{i\mathds{X}^{(Gluon Bundle)}}e^{i\mathds{X}^{(Quark Loop Chain)}}$
\begin{equation}
    T(s,\vec{q})= \frac{is}{2M^2}\int d^b e^{i\vec{q}\cdot \vec{b}}[1- e^{\sqrt{i}g\delta^2_q\varphi(\vec{b})/2}e^{g^2\delta_q^2(\kappa/\bar{m}^2)\Delta\bar{\varphi}/4}]
\end{equation}

Expanding above we get for one gluon bundle plus one quark loop chain exchange:
\begin{dmath}
    T_1(s,\vec{q})=\frac{s}{2M^2}\frac{g}{2}(\frac{\lambda}{m})^2(\frac{m}{E})^{2p}[-\frac{1}{\sqrt{2}}e^{-q^2/4m^2}+i(\frac{1}{\sqrt{2}}e^{-q^2/4m^2}+\frac{g}{2}\kappa\frac{q^2}{\bar{m}^2}e^{-q^2/2\bar{m}^2})]
\end{dmath}
giving a differential cross section, (green dotted curve in figures)
\begin{dmath}
    \frac{d\sigma_1}{dt}(s,q^2) = K \frac{27}{4\pi}\frac{g^2}{4}(\frac{\lambda}{4})^4(\frac{6m}{\sqrt{s}})^{4p}[\frac{1}{2} e^{-q^2/2m^2}+(\frac{1}{\sqrt{2}}e^{-q^2/4m^2}+\frac{g}{2}\kappa\frac{q^2}{\bar{m}^2}e^{-q^2/2\bar{m}^2})^2]
\end{dmath}

And for one gluon bundle plus two gluon bundles plus one quark loop chain exchange:
\begin{dmath}
    T_2(s,\vec{q})=\frac{s}{2M^2}\frac{g}{2}(\frac{\lambda}{m})^2(\frac{m}{E})^{2p}[(-\frac{1}{\sqrt{2}}e^{-q^2/4m^2}+\frac{1}{2}\frac{g}{2}\delta^2_q\frac{m^2}{2\pi}e^{-q^2/8m^2})+i(\frac{1}{\sqrt{2}}e^{-q^2/4m^2}+\frac{g}{2}\kappa\frac{q^2}{\bar{m}^2}e^{-q^2/2\bar{m}^2})]   
\end{dmath}

with differential cross section (red solid line in figures)
\begin{dmath}
    \frac{d\sigma_2}{dt}(s,q^2)=K \frac{27}{4\pi}\frac{g^2}{4}(\frac{\lambda}{m})^4(\frac{6m}{\sqrt{s}})^{4p}[(\frac{1}{2} e^{-q^2/2m^2}-\frac{g}{2}\frac{\lambda^2}{4\pi}(\frac{6m}{\sqrt{s}})^{2p}e^{-q^2/8m^2})^2+(\frac{1}{\sqrt{2}}e^{-q^2/4m^2}+\frac{g}{2}\kappa\frac{q^2}{\bar{m}^2}e^{-q^2/2\bar{m}^2})^2]
\end{dmath}

$K$ is conversion factor from mb to GeV at $0.44 mb GeV^{-2}$.
For the ISR data, we obtained parameters:
g=7.0, p=0.13, $\lambda=0.5$, $\kappa=-6.8 10^{-4}$, $m=0.23GeV \approx 1.5 m_{\pi}$, $\bar{m}=0.64 GeV \approx 4.5 m_{\pi}$.

For the LHC TOTEM data, we obtained 
$g=7.0$, $p=0.55$, $\lambda=0.72$, $\kappa = -4.2 10^{-3}$, $m=0.16 GeV \approx m_{\pi}$, $\bar{m}=0.41 GeV \approx 3 m_{\pi}$.

\section*{Acknowledgments}

This work was made possible by a generous grant from the Julian Schwinger Foundation.

\section*{References}

%\section*{References}

\end{document}